# Machine learning aided materials design platform for predicting the mechanical properties of Na-ion solid-state electrolytes

Junho Jo, Eunseong Choi, Minseon Kim, and Kyoungmin Min [*]

School of Mechanical Engineering, Soongsil University, 369 Sangdo-ro, Sangdo-dong, Dongjak-gu, Seoul 06978, Republic of Korea

## Abstract

Na-ion solid-state electrolytes (Na-SSE) exhibit high potential for electrical energy storage owing to their high energy densities and low manufacturing cost. However, their mechanical properties critical to maintain structural stability at the interface are still insufficiently understood. In this study, a machine learning based regression model was developed for predicting the mechanical properties of Na-SSEs. As a training set, 12,361 materials were obtained from a well-known materials database (Materials Project) and were represented with their respective chemical and structural descriptors. The developed surrogate model exhibited a remarkable accuracy ($R^2$ score) of 0.72 and 0.87, with a mean absolute error of 11.8 GPa and 15.3 GPa for the shear and bulk modulus, respectively. This model was then applied to predict the mechanical properties of 2,432 Na-SSEs, the properties of which have been validated with first principles calculations. Finally, the optimization process was performed to develop an ideal materials screening platform by adding the new minimized dataset, wherein the prediction uncertainty is reduced. We believe that the platform proposed in this study can accelerate the search for Na-SSEs with ideal mechanical properties at minimum cost.

[*]Corresponding author: kmin.min@ssu.ac.kr (K. Min)

# INTRODUCTION

Due to the increasing demand for electrical devices, such as electric vehicles, energy storage systems, and mobile phones, it is highly imperative to develop batteries with high energy densities and long-term cyclability. In this regard, Li-ion batteries have been successfully commercialized and continue to progress in terms of capacity and reliability. However, the lack of Li reserves has prompted the search for their alternatives. In this regard, the research on Na-ion batteries as proposed alternatives to Li-ion batteries is rapidly growing due to the abundance of Na sources. [1]–[9]

The electrolytes implemented in Na-ion batteries are primarily in the liquid-state due to their high ionic conductivity. However, the possibility of leakage during the battery operation coupled with their intrinsic flammability could lead to catastrophic battery failures. For these reasons, all-solid-state sodium batteries (ASSB) have attracted significant attention because of their structural stability and high energy density. [10] In addition, the capacity of ASSBs could be increased by the deployment of solid-state electrolytes (SSE) that could prevent the short circuit resulting from the dendritic growth occurring on the Na metal anode. Despite their advantages, the application of ASSBs is confronted with several challenges, such as high impedance, poor contact stability, and low ionic conductivity. [11] Furthermore, the mechanical properties of SSEs have not yet been fully examined, which is critical for blocking dendrite penetration and maintaining interfacial stability.

For the SSEs to meet the above-mentioned criteria, the two most important mechanical properties are (1) the shear modulus, for limiting the dendritic growth, and (2) the bulk modulus for maintaining a stable interfacial contact between the electrolyte and the electrode. [12] Therefore, the search for the optimal electrolyte material necessitates having either a sufficient number of databases on mechanical properties or a surrogate model for predicting them. However, several previously reported studies have primarily focused on improving either the interfacial resistance or the electrochemical properties of the SSEs. Notable examples include reports on materials available as electrolytes and electrodes due to the high ionic conductivity and structural stability of NASICON materials, [13] ionic conductivity and thermoplastic processability in a series of Na-rich antiperovskites, [14] and high throughput screening of ionic conductivity of solid electrolytes [15].

It is also important to note that one of the largest materials databases for mechanical properties can be found from the Materials Project (MP), which has a calculated elasticity database of 12,361 inorganic materials. [16] However, only a few Na-containing compounds are calculated in the MP database; only 442 (6.48%) structures out of the 6817 possible Na-SSE candidates are known. To overcome this data shortage, performing additional density functional calculations (DFT) is essential. [17] However, owing to the high computational cost, it is practically inefficient to compute the remaining unexplored structures.

To overcome these issues, several studies related to materials screening implement machine learning and deep learning algorithms instead of performing brute-force experiments or calculations for all of the search spaces, which have proven to be outstanding in terms of accuracy and efficiency. [18] The most relevant studies in this regard include, 1) the bandgap prediction model of inorganic solids by machine learning (solely with the materials compositions), [19] 2) the band gap prediction model for inorganic compounds by machine learning (confirmed with DFT calculations), [20] 3) exploiting the electronic structure using a machine learning model to replace the DFT calculations, [21] 4) prediction of the phosphor host crystal structure's Debye temperature with support from the vector machine regression model, [22] 5) and the lattice thermal conductivity prediction of inorganic materials using a Gaussian process regression algorithm [23].

In this study, a materials screening platform for predicting the mechanical properties of Na-SSEs was developed with the aid of machine learning and high-throughput computations. For constructing the initial database for training, the elasticity information from the MP was obtained and represented with chemical and structural descriptors along with additional materials properties. The machine learning regression model was then constructed and implemented for predicting the shear and bulk modulus of potential Na-SSEs. Additionally, to minimize the prediction uncertainty in the trained model, the optimization process was employed and its performance was evaluated. For validating the functionality and accuracy of the current platform, high-throughput DFT calculations were performed for the potential Na-SSEs candidates.

**METHODS**

**Figure 1** illustrates all the processes undertaken in this study. The first sequence involves constructing a promising database from the MP and descriptor set for our initial database. It should be noted that for the elastic properties, i.e., the shear ($G_{VRH}$) and bulk modulus ($K_{VRH}$), the Voigt-Reuss-Hill (VRH) approximation is employed for the elastic properties, i.e., the shear ($G_{VRH}$) and bulk modulus ($K_{VRH}$) which are predicted for screening the elasticity of the Na-SSE candidates. [24] After constructing the initial database, the input features are preprocessed with scaling and the removal of outliers. Then, the regression model is constructed by comparing several sampling methods by modifying the distribution of the training database (for example, in stratified sampling vs. including structures, the mechanical properties of which are larger than a certain value in the training set). Thereafter, this surrogate model was used to predict the unexplored mechanical properties of 2,432 Na-containing materials.

Moreover, we performed DFT calculations for the $G_{VRH}$ and $K_{VRH}$ of these materials. However, we did not apply these calculations in the model constructed from the initial database to evaluate the effectiveness of the optimization process as this will be discussed later in this study. The performance of the initial model was evaluated by calculating the associated uncertainty and prediction accuracy. The exploration method involved the quantification of uncertainty of the model in the standard deviation (STDEV) and its subsequent use in the optimization process. We added data from the DFT database to the training sets during the optimization process and compared its performance with the random addition for validation.

**Machine Learning and Feature Engineering**

We constructed a database that was widely used and verified for material design. The mechanical properties, $G_{VRH}$ and $K_{VRH}$, of 12,361 materials along with 8 properties (volume, density, bandgap, formation energy per atom, energy above hull, number of atoms, number of elements, crystal system number) were downloaded from the MP database to be used as the initial database. For constructing machine learning features, the descriptor set consists of chemical [25] and structural descriptors [26]. Chemical descriptors are available in a total of 145 features consisting of elemental properties, which is based on fundamental atomic properties, such as electronegativity, melting point, and so on. This set was used as descriptors for predicting the electronic properties of crystalline compounds and glass formability of metallic alloys with high accuracy. [27], [28] Meanwhile, the structural descriptors consist of 126 features derived from the Voronoi tessellation based on the compound's crystal structure.[26], [19] This descriptor set proved excellent in predicting the structural characteristics in the crystal system and space group prediction model [29] and in predicting the formation energy of inorganic solid materials [30].

The machine learning regression algorithm is based on the gradient boosting method, LightGBM, which has been chosen because of its proven performance in many machine learning projects. It is able to process large amounts of data and is computationally efficient as well. [31] Thereafter, the final prediction model is constructed by adopting the descriptors with high feature importance and the sampling type exhibiting high prediction accuracy. After that, the Na-containing materials, the mechanical properties of which are yet unknown in the MP database, are predicted through the constructed model and the uncertainty of the model is quantified with STDEV value. Finally, the predicted values are compared with the calculated ones used to indicate the prediction accuracy of the model and the STDEV guides the model's optimization process.

**Na SSE candidates and model optimization**

In addition to the 12,361 datasets used as the initial database, the $G_{VRH}$ and $K_{VRH}$ values of 2,432 SSE candidates were obtained through the DFT calculations. These data are described with the same descriptor sets used in initial database. Detailed parameters for the DFT calculations to obtain the elasticity information are adopted from our previous study. [8]

Active learning optimizes the prediction model by adding selected databases to the training sets which are known to be excellent in reducing the computational cost, while improving the model performance. Optimization techniques has been employed in several research studies and it has been helpful in 1) finding the optimal bandgap and refractive index with minimized resources, [32] 2) Bayesian optimization to find the optimal optoelectronic and thermoelectric materials, [33] 3) the discovery of low thermal hysteresis NiTi-based shape memory alloys using inference and global optimization, [34] 4) surrogate-based analysis and optimization strategy for designing liquid rocket injectors, and [35] 5) developing an open-source python library for Bayesian optimization library and application examples for determining the atomic structure of a crystalline interface. [36]

In this regard, the exploration method is employed for improving the prediction accuracy and reducing the uncertainty. [32] The prediction model produces 20 different predicted values for each material as it is trained on 20 different data sets according to the random states used for predicting the unknown data. Thus, STDEV of these predicted values is used to quantify the prediction uncertainty. Then the materials with high STDEV values are included in the training set and the surrogate model is trained again. This optimization process is repeated until 2,400 out of the 2,432 (all of the Na-SSE candidates) data sets are included in the training set. Additionally, its performance is compared with random data addition for validating the optimization. We evaluate the advantage of this method by recording the changes in the coefficient of determination ($R^2$), mean absolute error (MAE) of the model, and the STDEV of the MAE of the remaining database during active learning.

**RESULTS AND DISCUSSIONS**

**Database construction**

For constructing the training database, we used 12,361 materials. This number of materials was obtained as follows. First, we extract 13,116 materials, the $G_{VRH}$ and $K_{VRH}$ values of which are calculated and available in the MP database, and then obtain their corresponding properties. If the material is found unsuitable for use as an electrolyte or the reliability of the calculated values is suspected due to the corresponding scales of the properties involved, it could eventually affect the reliability of the database, and thus, is removed if 1) $G_{VRH}$ or $K_{VRH}$

< 0 GPa, 2) $G_{VRH}$ > 700 GPa and $K_{VRH}$ > 500 GPa, 3), the material contains an inert gas, 4) and the material contains radioactive elements.

Based on the above criteria, the number of materials is finally reduced to 12,361, and their distribution is shown in **Figure 2(a)**. The substances with $G_{VRH}$ and $K_{VRH}$ values of 0–100 GPa and 0–200 GPa accounted for 90.0% and 89.8% of the total, respectively. Among the 12,361 data sets, only 442 Na-containing compounds are available, and their distribution is shown in **Figure 2(b)**. On the other hand, the $G_{VRH}$ and $K_{VRH}$ values of 0–60 GPa and 0–100 GPa accounted for 89.8% and 88.2% of the total data, respectively. Since the entire database is concentrated on relatively low property values and most of the Na compounds seem to have relatively lower mechanical properties than other material groups, it is expected that the data with high mechanical properties should be included in the training set so that predicting the Na-SSE candidates with high mechanical characteristics could be possible.

A total of 6,807 potential Na-SSE materials are available in the MP database. However, prior to performing the DFT calculations, some of the materials are screened out because they do not satisfy the fundamental criteria, i.e., they 1) contain an inert gas, 2) are structurally unstable and unable to make a structural descriptor, 3) have a band gap < 0.1 eV, 4) a formation energy > 0.0 eV/atom, 5) a convex hull of energy > 0.1 eV/atom, 6) the total number of atoms in the structure is ≥ 50, 7) the formation energy is larger than its polymorph structures, 8) contains an electrochemically active element, 9) and does not converge with the DFT calculations. These criteria finally reduced the number of structures to 2,432.

The following process was carried out to derive the optimal descriptor setting. The composition of the descriptor was differently composed of All (8 + 145 + 126), C + S (145+126), C (only chemical descriptor set), and S (only structural descriptor set) descriptor sets, and their prediction accuracy was compared. Of the total database, 80% was used as the training set, while the remaining 20% was used as the testing set. As shown in **Figure 3(a)**, the best prediction accuracy was obtained when 8 descriptors imported from the MP were additionally configured with the structural and chemical descriptors, respectively. Therefore, all descriptors were viable for model training. However, for the cost-effective descriptor setting of the model, the performance comparison was carried out after removing the convex hull energy ($E_{hull}$) and formation energy ($E_{form}$) of the descriptors that cause additional calculations from the descriptor set. As shown in **Figure S1 (Supporting Information (SI))**, the $R^2$ decreased slightly from 0.734 to 0.720 for $G_{VRH}$ and from 0.877 to 0.874 for $K_{VRH}$. These two descriptors are excluded from rest of the study as they hardly affect the model performance.

In addition, the feature importance of all the descriptors is obtained (provided with a separate file in **Table S1 (SI)**) and the four most important descriptors among them are shown in **Figure 3(b)**. There are three descriptors that are directly obtained from the MP database (volume, density, and number of atoms). This indicates that the MP descriptor is effectively used in the model. Moreover, we observed that the structure-related descriptors are effectively applied with them to predict the mechanical properties (**Table S1 (SI)**). Of the 25 most important features, the structure-related descriptors are 12 and 9 each for $G_{VRH}$ and $K_{VRH}$, respectively. From this we could reconfirm that the features from the MP and the structural and chemical descriptors have been appropriately used in the model

**Sampling strategy**

Since the training and testing set configuration can significantly affect the prediction accuracy, it is imperative to investigate which of these datasets should be included to fulfil this requirement. In this regard, three different sampling methods are applied depending on the materials composition and how the samples are selected. *Model 1* is constructed using a random sampling method wherein Na materials are included only in the test set. *Model 2* is constructed using a random sampling method and Na materials are included both in the training and the testing set. *Model 3* is constructed using a stratified sampling method, wherein the materials with $G_{VRH} \geq 120$ GPa are included in the training set for the shear modulus prediction, and the materials with $K_{VRH} >i\ 350$ GPa are included in the training set for the bulk modulus prediction. The criterion in *Model 3* is set by analyzing the prediction accuracy variation with respect to the change in the minimum value to be included, as shown in **Figure S3 (SI)** in details.

As shown in **Figure 4**, the prediction accuracy of *Model 2* is higher than that of *Model 1*. This means that the prediction model is improved when the Na materials are included in the training set which could be because the Na-containing materials are not satisfactorily predicted in *Model 1*, as illustrated in **Figure S2 (SI)**. Although the $G_{VRH}$ of Na-containing materials showed a high prediction accuracy ($R^2$ of 0.752), their STDEV was 6 times larger than that of other materials. Similarly, $K_{VRH}$ also showed the highest $R^2$ of 0.894, but the corresponding STDEV was more than twice that of other materials. This indicates that in order to build an effective mechanical properties prediction model for Na-SSE candidates, the Na-containing materials should be included in the training set. In the case of *Model 3*, although stratified sampling generally yields a better model than random sampling, we increase the prediction accuracy additionally by including materials that have properties exceeding above a certain criterion in the training set. As a result, we developed a better prediction model than *Model 2* (**Figure 4**). In the case of $G_{VRH}$, the selection criterion was to include materials with a $G_{VRH} \geq 200$ GPa in the training set due to their highest accuracy and a low uncertainty. In the case of $K_{VRH}$, the highest prediction accuracy was obtained for $K_{VRH} \geq 450$. Although the difference was small, the STDEV was large, and thus, the criterion was modified to include materials with $K_{VRH} \geq 350$ as a training set. Finally, with the *Model 3*, 2,432 unexplored Na-SSE candidates' data was predicted with an $R^2$ of 0.720 (STDEV: 0.016) for $G_{VRH}$, and 0.874 (STDEV:0.006) for $K_{VRH}$.

**Uncertainty quantification**

Although the constructed model is able to make moderately accurate predictions, the STDEV in the mechanical properties of 2,432 Na SSE candidates are also obtained to quantify the uncertainty in *Model* 3, as shown in **Figure 5(a)**. In the case of $G_{VRH}$, 89% of the predicted values with a STDEV ≤ 6 GPa are obtained, while for $K_{VRH}$, 93% of the predicted values with a STDEV ≤ 10 GPa are obtained. Most materials have a low STDEV, implying that the model prediction uncertainty is not significant. **Figure 5(b)** shows a scatter plot comparing the values predicted by the model and the values obtained from the DFT calculations. The MAE value of the model for $G_{VRH}$ and $K_{VRH}$ is 8.76 GPa (STDEV: 0.268 GPa) and 13.4 GPa (STDEV: 0.256 GPa), respectively. Moreover, it can be reconfirmed that the predicted values of most materials are similar to the calculated values as shown in **Table S1 (SI)**.

**Active learning**

Despite the model's low uncertainty its ability to predict materials with properties comparable to the ones in the initial database, there are some materials that are not well predicted. Moreover, this trend is more frequent for materials calculated to have high mechanical properties, which can be attributed to insufficient data on high mechanical properties yet. At this time, if the material database can be additionally configured through the optimized material addition, the prediction accuracy of the model can be further increased and the uncertainty can be critically reduced. In this aspect, the exploration method is implemented by adding the datasets having larger uncertainty than others into the training sets. Each optimization step performed involves 1) calculating the STDEV on the prediction of the mechanical properties, 2) adding the top 5 substances among the calculated STDEV to the model training set, 3) and iterating until 2,400 substances are added to the training set. We also investigated how the number of added data sets affects the optimization performance (5, 25, 50, 100, 200), as shown in **Figure S4 (SI)**. This indicates that the number of added data sets has almost no effect on the MAE trend change (for the remaining data). Therefore, an optimization process that adds 5 data sets to each iteration is taken as a representative in this study. We also create a comparative group to demonstrate the optimization performance. This is performed by randomly adding 100 data sets to the training set.

**Figure 6(a)** shows that using the exploration method improves the prediction accuracy much more significantly in comparison to the random case. As the iteration of the exploration method progresses, the MAE continues to decrease, which also decreases its STDEV. In other words, the prediction accuracy of the model continues to improve when the uncertainty decreases. Although adding the random data decreases the MAE and improves the prediction accuracy of the model, it is considerably less efficient, and the STDEV becomes even larger compared to the one obtained from the exploration method. This implies that data addition can generally improve the prediction performance but is highly inefficient. More specifically, the model performance of adding 1000 data sets with the exploration method is better than adding over 2,000 random data sets wherein the prediction accuracy (adding 2000 data) has a MAE of 5.27 GPa (STDEV 0.256 GPa) and 8.13 GPa (STDEV 0.651 GPa).for $G_{VRH}$ and $K_{VRH}$, respectively.

**In Figure 7**, the comparison between the predicted and actual values is shown when 1000 calculated materials are included as a training set through the optimization process. The number of predicted data sets is 1,433, which is 1,000 less than the scatter plot shown in **Figure 6**, and this clearly shows significant reduction in the difference between the predicted and calculated values in majority of the data. These results indicate that the active learning makes it possible to efficiently find and calculate the data that is not well predicted, while simultaneously increase the prediction accuracy rapidly, and decrease the model uncertainty. In conclusion, this optimization process eliminates the need to perform complex and time-consuming DFT calculations on all the unknown data. All the predicted data post-optimization are listed in **Table S3 (SI)** along with the calculated values**.**

## CONCLUSIONS

In this study, we implemented a machine-learning algorithm on potential SSE candidates to screen the mechanical properties, $G_{VRH}$ and $K_{VRH}$. The model built with the sampling method took into account the distribution of the initial database while utilizing the well-known MP database and the descriptor set to yield $R^2$ of 0.720 and 0.874, and MAE of 11.8 GPa and 15.3 GPa (MAE) for prediction of $G_{VRH}$ and $K_{VRH}$, respectively. This model predicts the SSE candidates that have not yet been calculated by the MP. Compared to the DFT calculations, most materials were well predicted and had low uncertainty. In the case of the data predicted by this model, 163 and 228 substances were searched for $G_{VRH}$ and $K_{VRH}$, respectively, with the top 10% mechanical properties of the initial database present in the SSE candidate database.

We also optimized the model through the exploration method. The results obtained were surprisingly good with the possibility of preferentially selecting and calculating materials that were not well predicted. Moreover, adding 1,000 data sets to the model in an optimized method showed improved performance over 2,000 randomly added data sets. As a result, it was not necessary to perform extensive calculations using the ab-initio method because the model that underwent effective optimization exhibited good prediction accuracy for the remaining unknown materials. The model building technique and optimization method in this study are applied to the SSE candidate materials. Considering the effectiveness in predicting materials of which the mechanical properties are unknown, such an approach can be applied to other materials groups or chemical spaces.

**Acknowledgements**

This work was supported by the National Research Foundation of Korea (NRF) grant funded by the Korea government (MSIT) (No. No. 2020R1F1A1066519).

**Appendix A. Supporting Information**

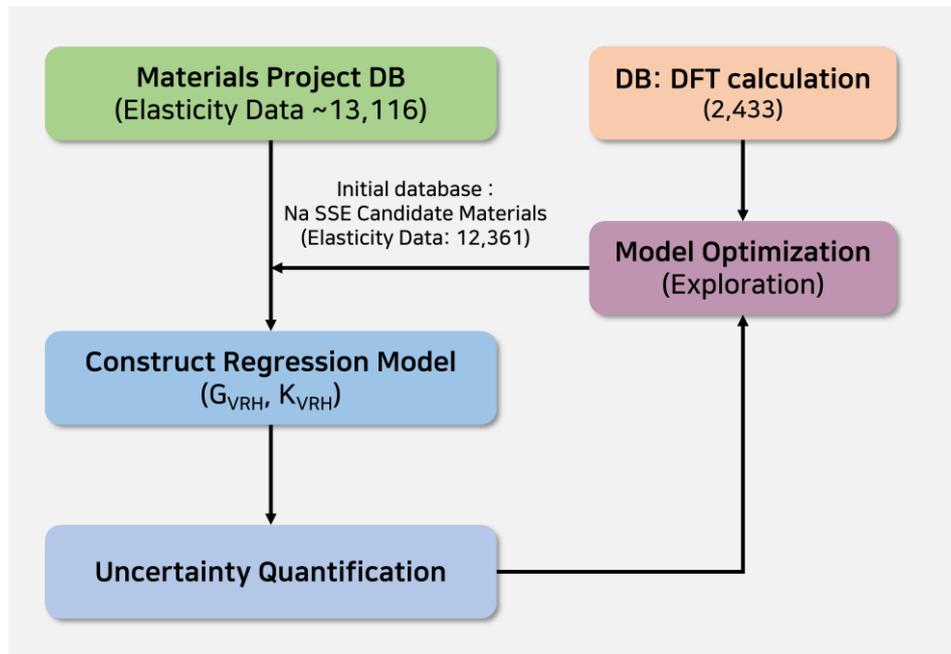

Figure 1. Flow chart of the machine learning and the optimization process for finding mechanically superior Na SSE candidates.

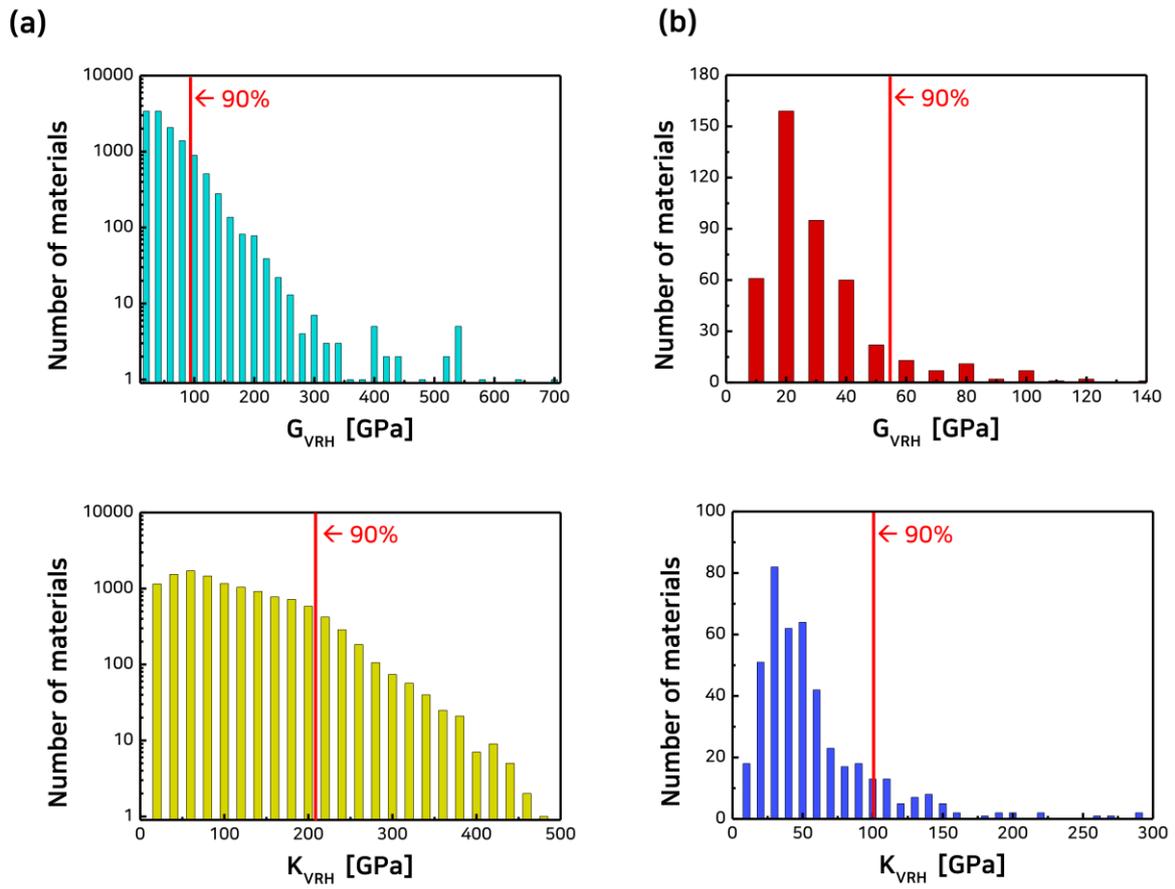

Figure 2. Initial database distribution of (a) all materials and (b) Na-containing materials for $G_{VRH}$ (top) and $K_{VRH}$ (bottom).

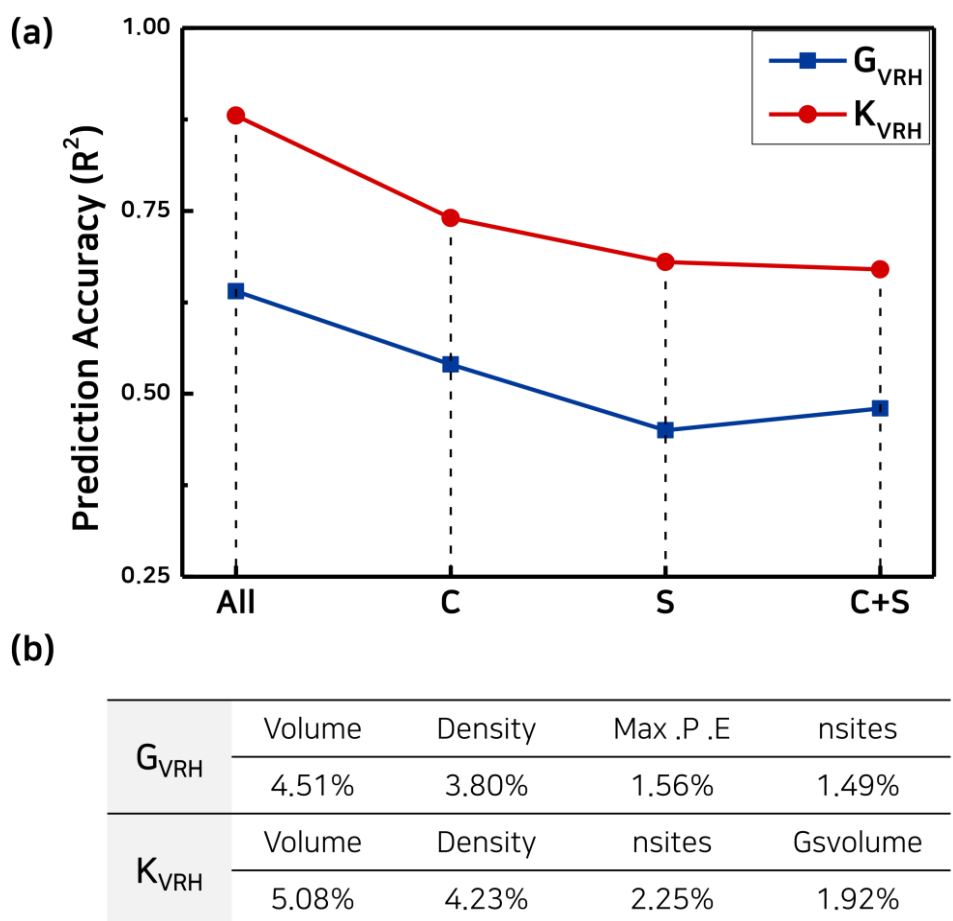

Figure 3. (a) Prediction accuracy change according to combinations of descriptor set. (b) Feature importance value of the most important four descriptors. (Max.P.E: maximum packing efficiency, nsites: numbers of atoms in compounds, GSvolume: volume per atom of ground state. M.P.E and Gsvolume are chemical descriptors and nsites is obtained from MP.)

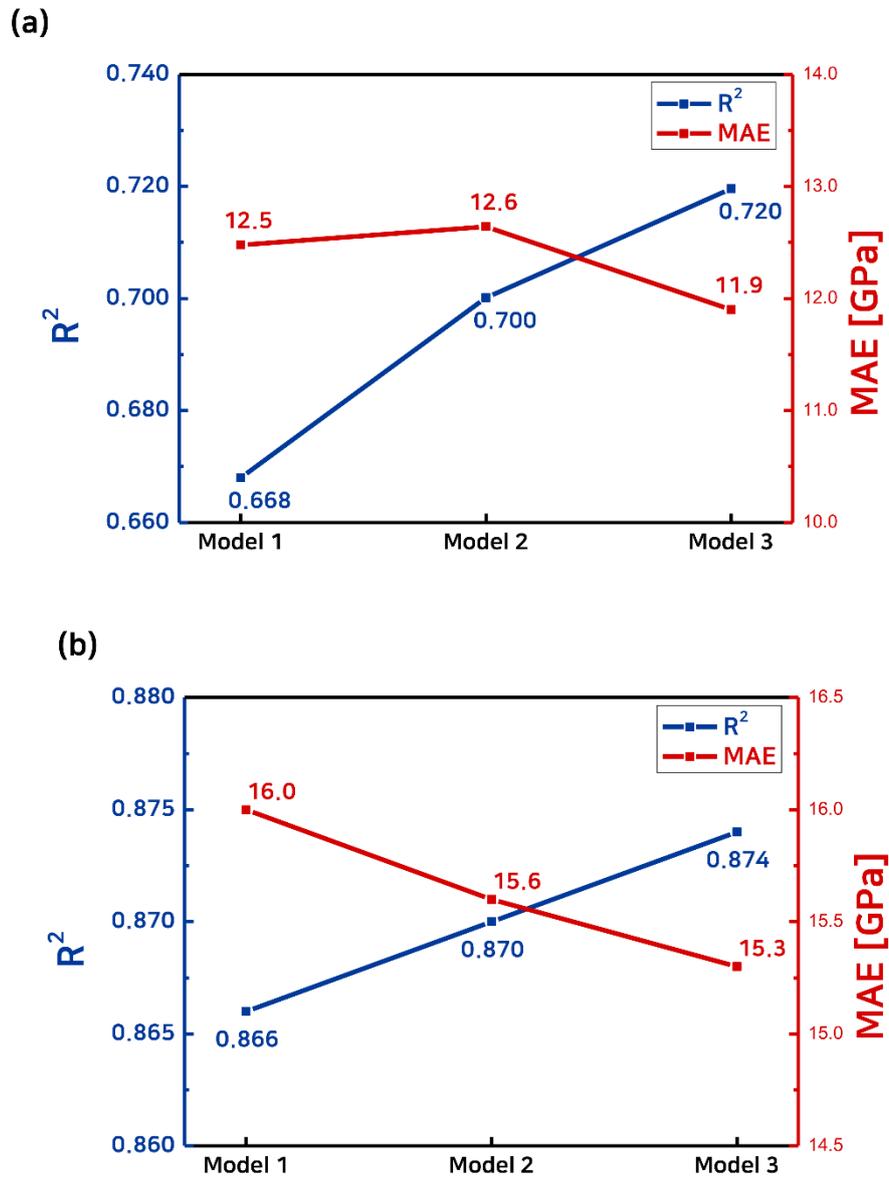

Figure 4. Prediction accuracy change according to different surrogate model with different composition of training sets for (a) $G_{VRH}$ and (b) $K_{VRH}$.

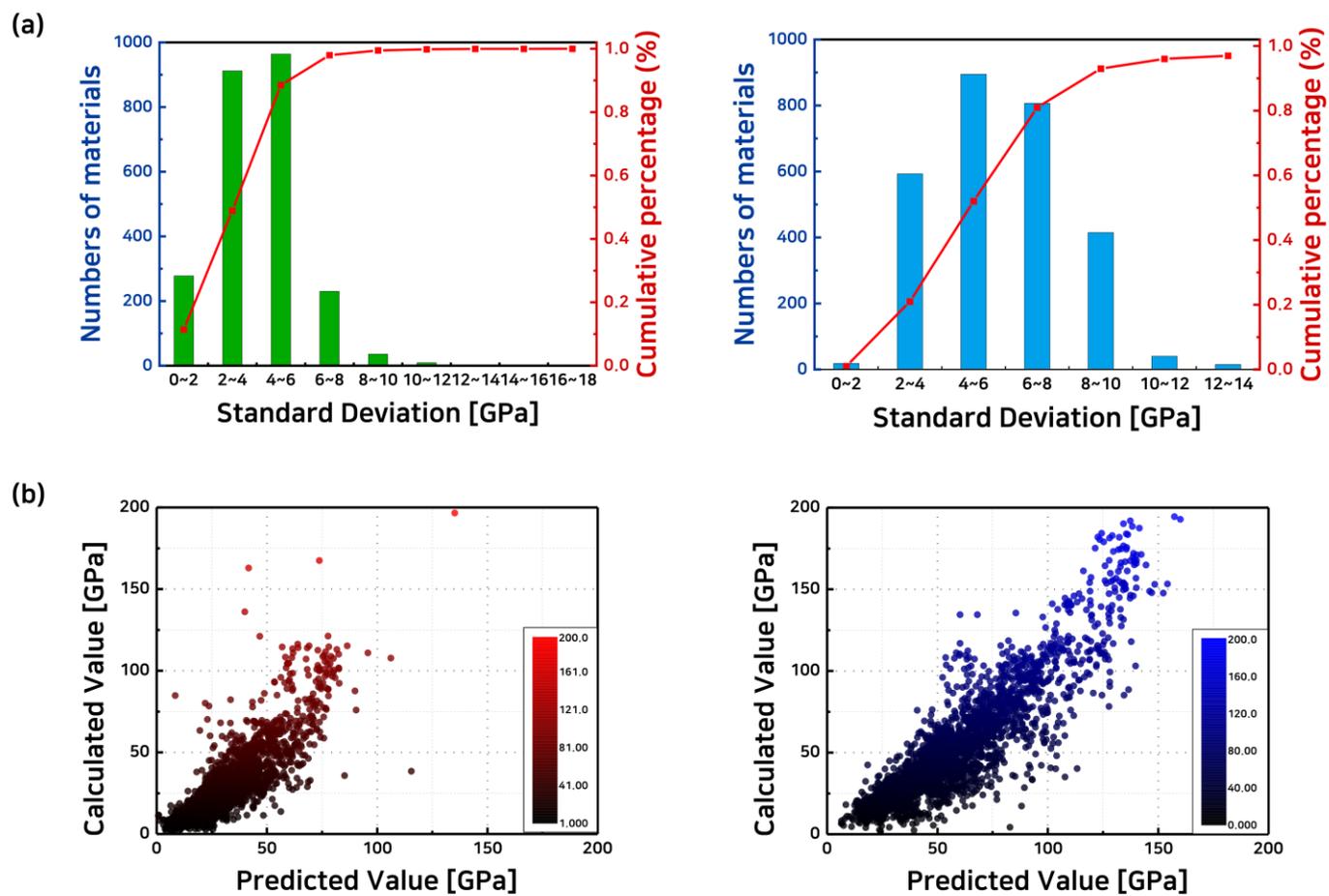

Figure 5. (a) Uncertainty analysis and its distribution and (b) predicted vs. calculated value plot before performing optimization for Na SSE candidates for $G_{VRH}$ (left) and $K_{VRH}$ (right).

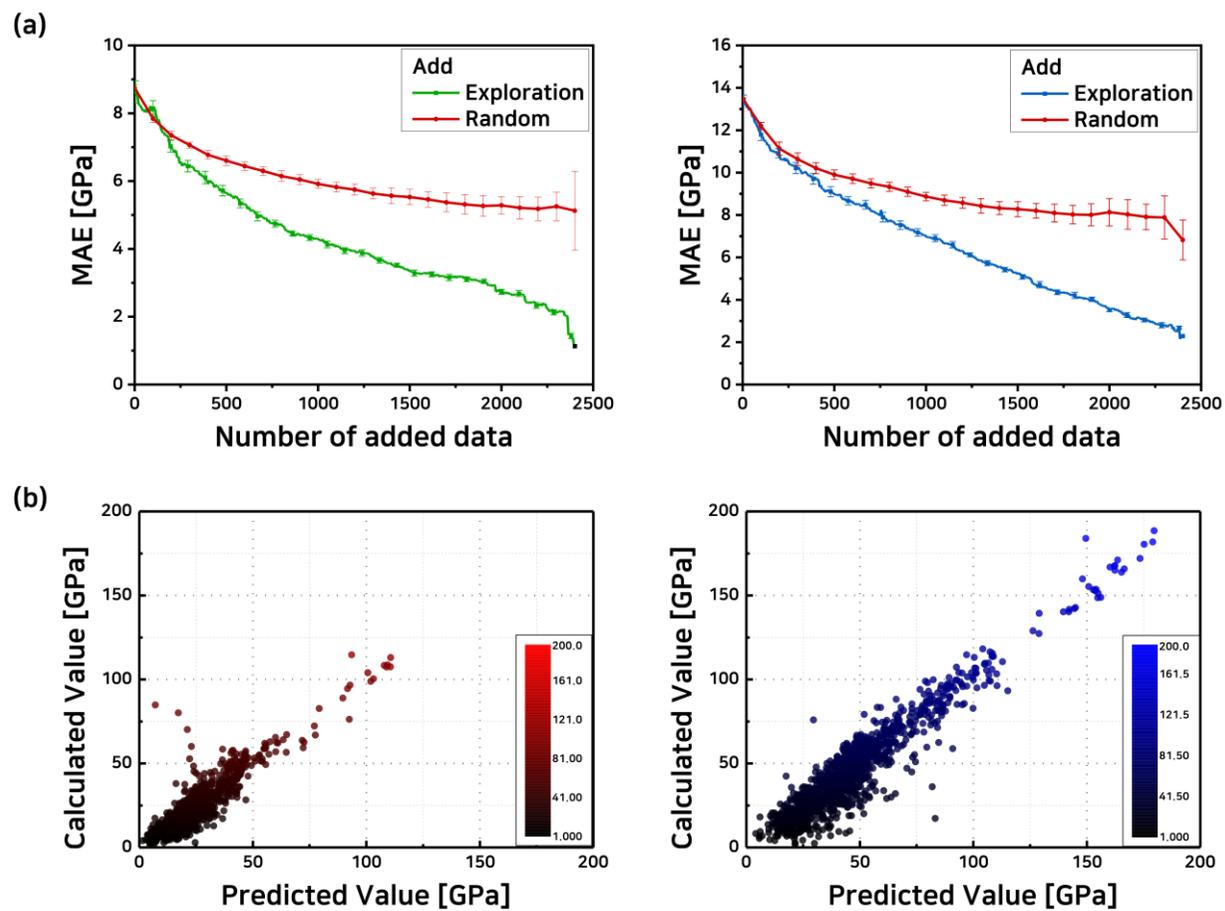

Figure. 6 (a) Prediction accuracy change during the optimization process and (b) predicted vs. calculated value plot after performing optimization for Na SSE candidates for $G_{VRH}$ (left) and $K_{VRH}$ (right).